\documentstyle[prl,aps,psfig,floats,twocolumn]{revtex}
\begin{document}
\draft
\twocolumn[\hsize\textwidth\columnwidth\hsize\csname
@twocolumnfalse\endcsname

\preprint{cond-mat/15-11}

\title{Surface Hardening and Self-Organized Fractality Through Etching of Random Solids}

\author{A. Gabrielli$^{1}$, A. Baldassarri$^{1,2}$, and B. Sapoval$^{1,2}$}

\address{$^{1}$Laboratoire de la Physique de la Matiere Condens\'{e}e,\\ Ecole
Polytechnique - CNRS, 91128 - Palaiseau, France\\
$^{2}$Centre de Math\'{e}matiques et de leurs Applications,\\
Ecole Normale Sup\'{e}rieure - CNRS, 94140 Cachan, France}

\date{\today }

\maketitle

\begin{abstract}

When a finite volume of etching solution is in contact with a
disordered solid, complex dynamics of the solid-solution interface
develop. 
If the etchant is consumed in the chemical reaction, the dynamics stop
spontaneously on a self-similar fractal surface.
As only the weakest sites are corroded, the solid surface gets progressively
harder and harder.
At the same time it becomes rougher and rougher uncovering the critical 
spatial correlations typical of percolation.
From this, the chemical process reveals the latent percolation criticality
hidden in any random system. Recently, a simple minimal model
has been introduced by Sapoval et al. to describe this phenomenon.
Through analytic and
numerical study, we obtain a detailed description of the
process. 
The time evolution of the solution corroding
power and of the distribution of resistance of surface
sites is studied in detail. This study explains the progressive hardening
of the solid surface.
Finally, this dynamical model appears to belong to the universality class of Gradient Percolation.
\end{abstract}
\narrowtext

\vskip2pc] {PACS numbers: 64.60Ak, 81.65Cf, 68.35Bs}

\section{Introduction}

Corrosion of solids has major economical consequences \cite{evans,uhlig}. 
It is also interesting from the point of view of theoretical physics of random
systems \cite{williams,nagatani,meakin,reigada,balazs2}. 

The comprehension of the basic physical mechanisms involved in corrosion
implies the study of the dynamical
evolution of the corrosion process and that of the morphological features of the 
corroded surface.

This paper presents a detailed study of a minimal model inspired
by recent experiments on pit corrosion of aluminum thin
films by an appropriate etching solution~\cite{exp-etc}.
This two dimensional model is a simplified etching model.
It was first introduced in~\cite{sapo-etch}, 
where a preliminary numerical study has been developed. 
It provides a simple description for the action
of a finite volume of a corroding solution on the surface of a
disordered solid.

When an etching solution is in contact with an initially flat surface of
a disordered solid, it starts to corrode its weakest regions and the surface 
gets ``harder''. However, at the same time,
new regions are discovered which contains weak elements.
Depending on the corrosion reaction mechanism,
different situations for this hardening process can occur.

Often the corrosive power of the solution is proportional to an etchant
concentration. 
If the etchant is consumed in the reaction, then the corrosive power of a 
{\it finite volume} of solution decreases during the time
evolution of the process. 
As the solid surface gets ``harder and harder'', and the corroding power of the solution
gets ``weaker and weaker'', the corrosion process stops spontaneously
in a finite time interval. At this moment all the surface sites are
``too hard'' to be etched by the solution.

It is this phenomenon which is studied both numerically and analytically in
this paper.

A most interesting aspect of this kind of dynamical corrosion is that
the final surface has a
fractal geometry, showing that the
corrosion mechanism itself uncovers the spatial correlations among the strong sites
belonging to the solid. This is why this phenomenon is intimately related 
to percolation properties of random systems.
In that sense this kind of corrosion reveals a ``latent'' criticality
embedded in any random system.

The model reproduces qualitatively the same phenomenology observed
experimentally \cite{exp-etc}. The dynamical evolution can be divided into two
different regimes:
\begin{enumerate}
\item In the first (smooth) regime, the corrosion is well directed and
the front becomes progressively rougher and rougher. 
In our model this regime does not depend on the details
of the discretization chosen, not even on the fundamental geometrical
features of the lattice, like the embedding space dimension or the
lattice coordination number.
\item In the second regime, the correlations revealed by the
hardening process become important: the dynamics becomes 
locally isotropic generating a fractal front. 
This corresponds to a critical regime, directly related to the
static percolation transition on the same lattice. 
\end{enumerate}

The hardness of the final interface, which is related to the final
corrosion power of the solution, depends on the external parameters as
the volume of the solution itself and the system size.
When the volume of the solution is not too large, one observes a geometrical scaling
regime.   
This regime corresponds to the scaling regime of a static percolation model known as ``Gradient
Percolation''. 
When the volume is increased, the correlation length grows to reach the system size.
Above this limit the finite size effects dominate the behavior, and we do not
study here this case.

\section{The model}
We first recall the two-dimensional etching model introduced in \cite{sapo-etch}.
Its schematic is shown in Fig.~\ref{fig0}:
\begin{itemize}
\item  The solid is represented as a site lattice (triangular or square), of
linear width $L$ and, eventually, infinite depth.
\item  A random number $r_{i}\in [0,1]$ (extracted from the flat probability
density function $\pi _{0}(r)=1$ for $r\in [0,1]$) is assigned to each solid
site $i$, representing its resistance to the etching by the solution. $r_{i}$
does not depend on time (quenched disorder), and on the site environment.
\item  The etching solution has a volume $V$ and is initially in
contact with the solid through the bottom boundary (see Fig.~\ref{fig0}).
It contains an initial number $N_{et}(0)$ of dissolved etchant molecules. 
\end{itemize}

Consequently, the initial concentration $C(0)$ of etchant in the
solution is given by: $C(0)=N_{et}(0)/V$. Calling
$N_{et}(t)$ the number of etchant molecules at time $t$,
$C(t)=N_{et}(t)/V$. At each time-step, the ``etching power'' of the
solution (i.e. the average ``force'' exerted by the solution on a
solid surface particle) is supposed to be proportional to $C(t)$ :
$p(t)=\Gamma C(t)$.
Hereafter the assumption $\Gamma =1$ is made, without loss of generality. It 
implies $C(t)\equiv p(t)$. At time-step $t$, all the interface
sites with $r_{i}<p(t)$ are dissolved and a
particle of etchant is consumed for each of these corroded solid sites.

\begin{figure}[tb]
\centerline{
\psfig{file=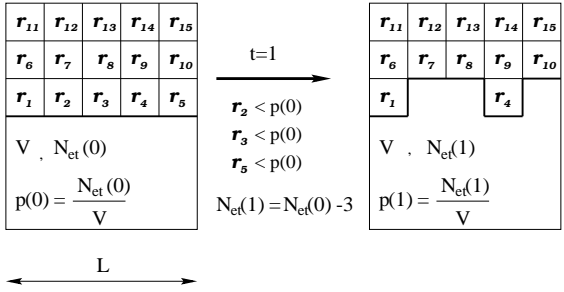,width=8cm}
}
\caption{
Sketch of the etching dynamics in a square lattice:
 the sites $2,3,5$ are etched
at the first time-step as their resistances are lower than $p(0)$.
Consequently the number of etchant particles in the solution decreases by $3$ 
units. At $t=1$, the new interface sites are $7,8,10$
if the solution can etch only the first nearest neighbors solid sites;
or the whole second 
layer if the solution can also etch the second nearest neighbors in a diagonal direction.}
\label{fig0}
\end{figure}

Let us call $n(t)$ the number of dissolved solid sites at time $t$. One can
express many important dynamical quantities through $n(t)$, or its time-integral 
$N(t)$, that is the total number of corroded solid sites up to time $t$.
The number of etchant particles in the liquid will decrease as: 
\begin{equation}
N_{et}(t+1)=N_{et}(t)-n(t)=N_{et}(0)-N(t)\,,  \label{eq1}
\end{equation}
and consequently the etching power of the solution is:
\begin{equation}
p(t+1)=p(t)-\frac{n(t)}{V}=p(0)-\frac{N(t)}V\,.  \label{eq2}
\end{equation}
Note that, as $p(t+1)<p(t)$, a site having resisted to etching at a certain time-step 
will resist forever. 
Consequently, the part of the solid surface which can be etched at time-step $t+1$
is restricted  to the sites which have been just uncovered by the etching process
at time $t$. We call this subset of surface the ``active'' part
of the surface. 
After a given time-step, all the solid sites which have been previously explored 
by the solution are definitely ``passive''.
However it may happen that 
``passive'' sites are disconnected from the bulk at a later time-step if they 
are connected to the solid by weak sites.

\subsection{Phenomenological description of the dynamics}

\begin{figure}[tbp]
\centerline{
\psfig{file=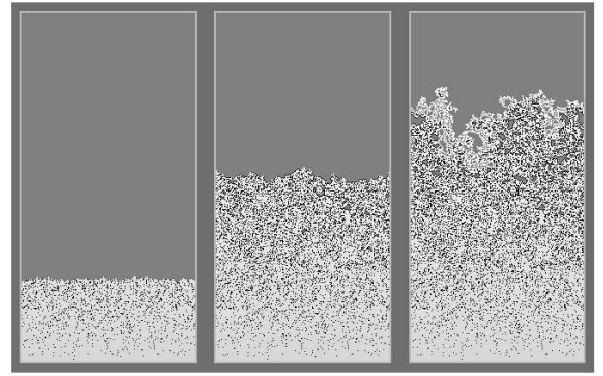,height=7cm,angle=-90}
}
\caption{Typical process represented at two intermediate
time steps, and at the final one. 
The solid is represented in grey, the solution in white, and the finite size
solid clusters detached by the solution in black.
The final solid surface is found to be fractal
up to a characteristic scale $\sigma$. 
}
\label{fig1}
\end{figure}

A typical process at two
intermediate times, and at the final time-step, is shown in Fig.~\ref{fig1}. 
Some finite solid clusters are detached
from the ``infinite'' solid by the corrosion process.
Consequently, at any time, the ``global
surface'' of the system is composed by both the finite clusters surfaces
and the surface of the infinite solid , which will be called the ``corrosion front''. 
Note that, in order to have a meaningful geometrical and physical definition
of the solution space and of the connected solid regions (and then of the
corrosion front), one has to use the so-called ``dual'' connection rules 
for solution and solid, respectively \cite{stauffer}.
For example, on the square lattice, if the solution etches both first and second nearest
neighbors, only first nearest neighbor solid sites should be considered as connected.
On the other side, if the solution etches only first nearest neighbor, both
first and second nearest neighbors solid sites should be considered as connected.
On the triangular lattice, if the solution etches first nearest neighbors, 
liquid and solid sites are considered to be connected
both by first nearest neighbors only. 

Two remarks should be made: 
\begin{itemize}
\item  
The corrosion front keeps quite smooth at the beginning of the dynamics 
(first snapshot of Fig.~\ref{fig1} ).
It becomes very irregular only towards the end of the dynamics (third snapshot of Fig.~\ref{fig1})
when $p(t)$ is close to the
percolation threshold $p_c$ on the same lattice \cite{sapo-etch}. 
\item The ``active'' part of the global surface is essentially restricted to the etching front,
since a site having resisted to the corrosion at a certain time-step will resist forever. 
\end{itemize}

\begin{figure}
\centerline{
\psfig{file=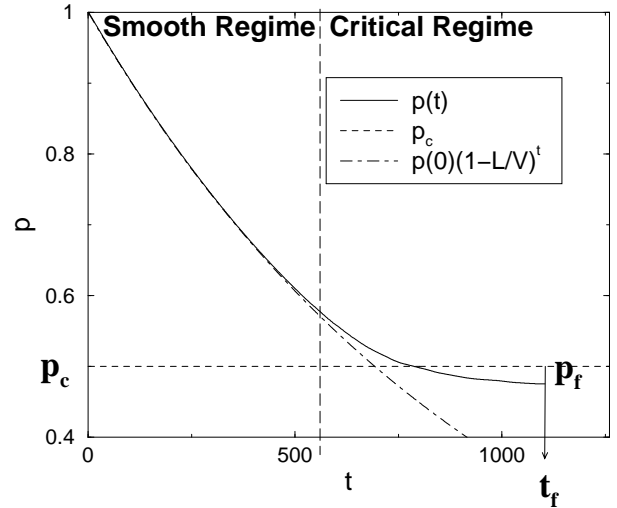,width=8cm}
}
\caption{
Decay of the corroding power $p(t)$ in a triangular
lattice with $p(0)=1$. The numerical evolution of $p(t)$
(solid line) is compared with the simple phenomenologic derivation of
Eq.~\protect\ref{simplest} (dashed line). When the two curves separate
the dynamics enters the ``Critical Regime'' dominated by percolation
effects.}
\label{dyn-fig}
\end{figure}

These observations are useful for a first analysis of the dynamics. 
Roughly speaking, if the front advances linearly, the
number of solid sites discovered at each time-step is
$L$ (the number of site in each layer).  Hence, in this
approximation, the number of etched sites is $n(t)=L\;p(t)$.
Using this approximation one gets (from Eq.~\ref{eq2}):
\begin{equation}
p(t)=p(t-1)\left(1-\frac LV\right)=p(0)\left(1-\frac LV\right)^t\,.
\label{simplest}
\end{equation}
This simple prediction is compared with the
actual simulation behavior of $p(t)$ in Fig.~\ref{dyn-fig}. 
The agreement between the simple
prediction~\ref{simplest} and the initial decay of $p(t)$ is very good
for values $p(t)> p_c$, i.e. in the {\em smooth regime} of the
dynamics. When $p(t)$ is close to $p_c$ this approximation is no more
valid and the dynamics enters the {\em critical regime}. 

A better derivation of Eq.~\ref{simplest}, and a more precise definition
of the two regimes will be given below providing a deeper insight on
the critical regime of the dynamics, when the surface becomes
fractal and the dynamics slow down and stop.

Note that the main hypothesis for the derivation
of Eq.~\ref{simplest} consists in assuming that at each time-step the number of new sites
checked for corrosion is always $L$, i.e. the whole next solid
layer. This is possible if the etching does not leave large connected
segments of uncorroded sites. In fact it is easy to show that the
non-etched sites, the number of which is approximatively $(1-p(t))L$,
are almost isolated, the average size of a segment of ``survived''
sites being
$\left<l\right>=1/p(t)$.

Interestingly the present phenomenological approach suggests an analogy
between our dynamical etching model and a static percolation model
known as Gradient Percolation \cite{sapo-etch,sapo2}.
This will be discussed next.

\subsection{Analogy with Gradient Percolation (GP)}

\begin{figure}[tbp]
\centerline{
\psfig{figure=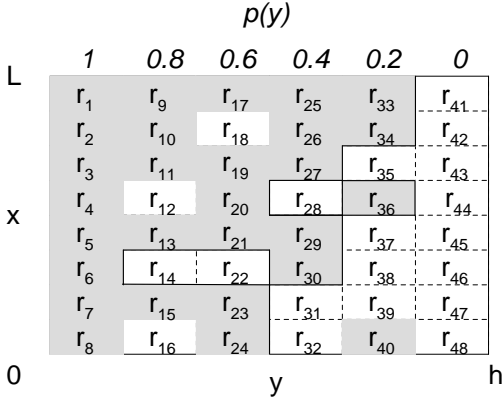,width=7cm}}
\caption{The Gradient Percolation Model. The numbers $r$ are chosen
randomly between $0$ and $1$. Each column has an occupation probability 
$p(y)$ ranging from $1$ at the left side to $0$ on the right side. 
A site $i$ is occupied if $r_i<p(y)$. The occupied and empty sites are represented respectively 
in grey and white. Apart from isolated islands and lakes, 
grey and white sites form two distinct connected regions.
The bold line represents the separation between these two regions.\label{gradperc}}
\end{figure}

The Gradient Percolation ({\bf GP}) problem \cite{sapo2,sapo3} can be
formulated in the following way: a random number $r_i\in [0,1]$ is assigned to each site of
a lattice of $x$-size $L$ and $y$-size $h$.  A constant
gradient of occupation probability in the $y$-direction is then imposed on the lattice: 
$p(y)=1-y/h=1-y\;\nabla p$.  The occupation rule is that
in each column the sites is occupied if and only if $r_i<p(y)$ (see
Fig.~\ref{gradperc}).
In the first column ($y=0$) the occupation probability is one, 
while in the last one ($y=h$) it is zero. 
These two special layers individuate two percolating clusters in the $x$ direction
of occupied (grey) and empty (white) sites (Fig.~\ref{gradperc}). The external frontier 
of the connected occupied cluster is called the {\em gradient percolation front} \cite{sapo2}. 
This front is centered around the layer with $p(y)$ equal to the
critical percolation threshold $p_c$ characteristic  of the lattice type.
The front is fractal with a dimension $D_f^{(GP)}\simeq 1.75$ up to a finite length
(front width $\sigma_{GP}$) which is a power law of the local gradient $\nabla
p=1/h$:
\begin{equation}
\sigma_{GP}\sim [\nabla p ]^{-\alpha_{\sigma}^{(GP)}}
\label{sigma-gp}
\end{equation}
where $\alpha_{\sigma}^{(GP)}\simeq 0.57$. 
Note that $ D_f^{(GP)}\simeq 7/4$ and $\alpha_{\sigma}^{(GP)}\simeq 1/D_f^{(GP)}$.
For this reason it was assumed that $D_f^{GP}$ is equal to the fractal dimension 
of the hull of the incipient infinite percolating cluster $D_f^h=7/4$ 
 in percolation theory \cite{duplantier,stauffer}.
The demonstration of the identity of the equivalent of $\alpha_{\sigma}^{(GP)}= 1/D_f^{(GP)}$
in percolation theory is given in \cite{duplantier}.

In addition, the occupation probabilities of the front range in an interval $p(y)\simeq p_c
\pm \Delta p$, where $\Delta p$ scales with the gradient as
\begin{equation}
\Delta p\sim [\nabla p ]^{\alpha_p^{(GP)}}\,.
\end{equation} 
The exponent $\alpha_p^{(GP)}$ is related to
$\alpha_{\sigma}^{(GP)}$, as $\Delta p\sim \sigma\,\nabla p$, which implies, from
Eq.~\ref{sigma-gp},
$\alpha_p^{(GP)}=1-\alpha_{\sigma}^{(GP)}$. 

Because of its characteristic properties, GP has provided
a powerful method to compute percolation threshold $p_c$ \cite{sapo-pc}.

In this model, one can
associate for each corroded site $(x,y)$ the value $p(x,y)$ of the solution etching
power at the time of corrosion of that site~\cite{sapo-etch}. In this way,
a position dependent ``field'' of occupation probabilities (by the solution) 
is spontaneously generated. This is the physical link with GP.
In the smooth time regime the successive active zones are consecutive solid layers containing
about $L$ sites. Consequently, $p(x,y)$ depends only on $y$ ($p(x,y)=p(y)$).

The ``active'' zone at time $t$ is then the whole layer at depth $y=t$
From Eq.~\ref{simplest} one can then write:
\begin{equation}
 p(y)=p(0)\left(1-\frac LV\right)^y\,.
\label{p-y}
\end{equation}
This equation defines a
{\em Self-Organized Gradient Percolation}, where the value of
the gradient depends on the parameter $L$ and $V$ as:
\begin{equation}
(\nabla p)_{et}\sim \frac LV
\label{gradient}
\end{equation}. 

\section{Simulations and numerical results}

Extensive simulations have been performed, considering triangular
and square lattices, with first nearest neighbour ({\bf f.n.n.}) and
second nearest neighbors  ({\bf s.n.n.}) (diagonal)
connections for the corrosion process.  All simulations start with 
$p_{0}=1> p_{c}$ in order to observe clearly 
the transition towards a critical regime, when $p(t)\simeq p_{c} $. 
Once $p_0$ is fixed, the parameter measuring the initial corroding ``force'' of the 
solution is $V=N_{et}(0)/p_0$. The other
fundamental parameter is the transversal size of the solid $L$. All
the data presented below refer to $1000$ different realizations of
the quenched disorder, for each choice of the parameters $L$ and $V$.

\subsection{Correlation length and ``Phase'' diagram}

In order to quantify the statistical properties developed by the
dynamical process, the average thickness of the final corrosion 
front is measured. 
If $\{y_{i}\}$ are the depths of the points $i$
belonging to the corrosion front at time $t$, its average thickness can be defined as: 
\[
\sigma =\sqrt{\frac{1}{I}\sum_{i=1}^{I}y_{i}^{2}-\left( \frac{1}{I}
\sum_{i=i,I}y_{i}\right) ^{2}} \,,
\]
where $I$ is the length of the corrosion front. 

\begin{figure}[tbp]
\centerline{
\psfig{file=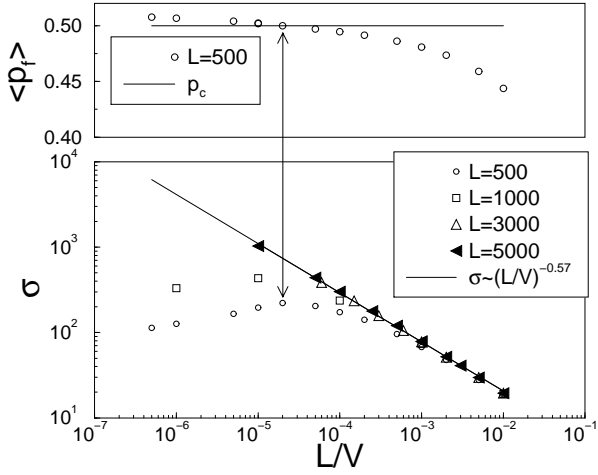,width=8cm}
}
\caption{Determination of the different corrosion behaviors. Bottom:
behavior of $\sigma$ as a function of $L/V$ for several sample sizes
(several values of $L$). Top: the behavior of $\left<p_f\right>$ for the smaller sample
($L=500$). Note that the maximum of $\sigma$ corresponds to the change of sign of
$\left<p_f\right>-p_c$ (vertical arrow).}
\label{sigma-fig}
\end{figure}

The behavior of the final value $\sigma$ at time $t_f$ 
as a function of the ``natural'' gradient
$L/V$ is shown in Fig.~\ref{sigma-fig} (bottom) for different fixed values of $L$. 
Several observations can be made:
\begin{itemize}
\item First, for sufficiently large values of $L/V$ 
(right side of Fig.~\ref{sigma-fig}) 
$\sigma$ follows the scaling behavior
\begin{equation} 
\sigma\sim (L/V)^{-\alpha_\sigma}
\label{sigma-eq}
\end{equation} 
with $\alpha_\sigma=0.57\pm 0.02$. This confirms the idea 
that the final features of our dynamical etching model, at least in this range of  $L/V$, 
are in the same universality class of GP once the identification $L/V\sim\nabla p$ is done, i.e. 
$\alpha_\sigma=\alpha_\sigma^{(GP)}$.
\item Decreasing $L/V$, $\sigma$ increases following the previous scaling behavior 
(Eq.~\ref{sigma-eq}) until reaching values of $L/V$ for which
$\sigma\simeq L$. For even smaller values of $L/V$,
a deviation from the aforementioned scaling law is observed. 
This deviation is characterized by a cross-over
to a region dominated by boundary effects. 
In this regime $\sigma$ seems to decrease slowly
together with the gradient $L/V$, instead of increasing.
\item Consequently, for a fixed value of $L$, one can distinguish a ``strong gradient'' process,
i.e. for values of $L/V$ in which Eq.~\ref{sigma-eq} holds, 
and  a ``weak gradient'' process for values of $L/V$ smaller than the cross-over value.
The cross-over between the two behaviors is marked by a marginal value of
$L/V$ for which $\sigma\simeq L$. Note that for this value
of $L/V$ the spatial correlations extends all over the sample.
Then this is a kind of ``critical''value of $L/V$.
\end{itemize}
Moreover one observes that $\left<p_f\right><p_c$ in the strong gradient regime
and $\left<p_f\right> >p_c$ in the weak gradient regime, where $\left<...\right>$
means an average over different realizations of the disorder with fixed parameters $L,V$. 
In this way the equality $\left<p_f\right>=p_c$ can be used
to identify the marginal (``critical'') value of $L/V$ for a fixed value of $L$. 

In the upper diagram of Fig.~\ref{sigma-fig},
the transition between the two regimes for $L=500$ is shown.
This transition corresponds to a value of $L/V \simeq 2\cdot 10^{-5}$ (marked by the 
double arrow crossing the two plots).

\begin{figure}[tbp]
\centerline{
\psfig{file=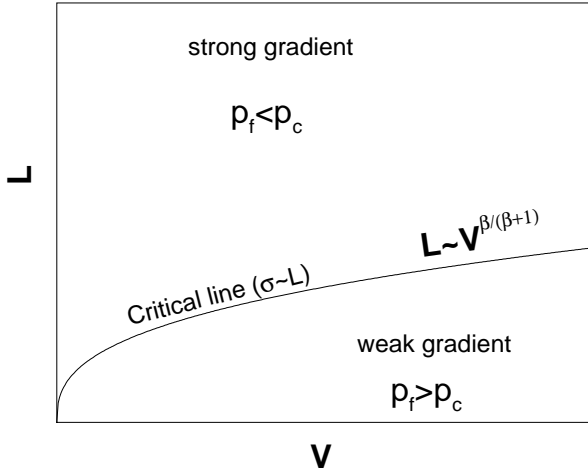,width=8cm}
}
\caption{``Phase'' diagram in the parameters space (see text).}
\label{phasediag}
\end{figure}

This behavior of $\sigma$ allows to sketch a kind of ``phase''
diagram for our model in the $(L,V)$ parameter space (Fig.~\ref{phasediag}).
The ``critical'' line $\sigma\simeq L$ separates the two ``phases''.
Here we use the terminology of phase transitions because in GP the correlation 
length is equal to the front width $\sigma$.

Since in the strong gradient ``phase'' Eq.~\ref{sigma-eq} holds, the  
scaling relation for the marginal line is:
\begin{equation}
L\sim V^{\frac{\alpha_\sigma}{1+\alpha_\sigma}}
\label{marginal-eq}
\end{equation}
The relevance of this relation with respect to the extensivity of spatial correlations
in the ``thermodynamic limit'' 
is discussed in Appendix A. 
In the following, we deal only with the strong gradient regime, 
leaving the detailed analysis of the weak gradient regime to further work.

\subsection{Strong gradient etching}

\begin{figure}[tbp]
\centerline{
\psfig{file=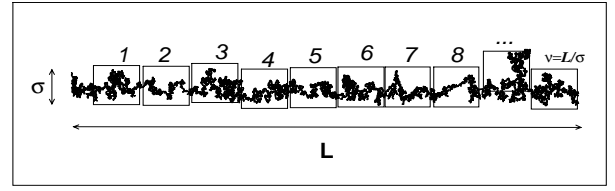,width=8cm,height=3cm,angle=0}
}
\caption{Typical final corrosion front for a simulation in
the strong gradient phase. Note that $\sigma\ll L$. Different non-overlapping almost 
independent regions are identified by numbers. The number $\nu=L/\sigma$ of
almost independent regions is relevant for the study of extremal
quantities \protect\cite{extremal}.
}
\label{interface}
\end{figure}

In order to study this regime, we focus on simulations of
sizes $L=3000$ and $L=5000$, with  $\left<p_f\right>\;<p_c$. 
Such values of $L$ are large enough, and at the same time they permit to collect 
large statistics. 
A typical corrosion front is presented in Fig.~\ref{interface},
where the condition $L\gg \sigma$ is emphasized. 
Note that on scales larger than $\sigma$, the corrosion front is almost flat.
This indicates the statistical independence among non-overlapping
regions of the surface of linear size larger than $\sigma$.

As mentioned earlier, $\sigma$ is described by Eq.~\ref{sigma-eq}.
Similarly to $\sigma$, other important properties
follow simple scaling relations with the gradient $L/V$ 
\cite{sapo-etch}.

\begin{figure}[]
\centerline{
\psfig{file=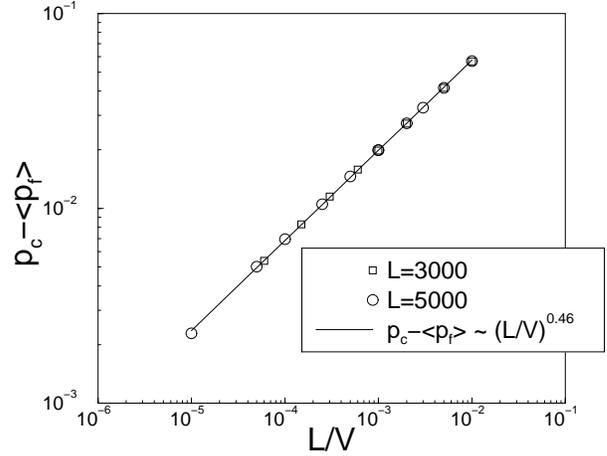,width=8cm}
}
\caption{Scaling behavior of $p_c-\left<p_f\right>$.
Note that, identifying $L/V$ with $\nabla p$ of
Gradient Percolation, one obtains the same values of the exponent
$\alpha_p=0.46\pm 0.02$.}
\label{fig2}
\end{figure}

The distance of the average value $\left<p_f\right>$ from $p_c$
follows the scaling law: 
\begin{equation}
p_c-\left<p_f\right> \sim \left(\frac{L}{V}\right)^{\alpha_p}\;,
\mbox{with}\; \alpha_p=0.46\pm 0.02  \label{delta-p}
\end{equation}
as shown in Fig.~\ref{fig2}.
\footnote{Appendix A discusses 
the possibility of obtaining a value of $\left<p_f\right>$ 
arbitrarily near $p_c$, remaining in the ``strong gradient'' 
region of Fig.~\ref{phasediag}.
Starting from a couple $(L_0,V_0)$ in the strong gradient phase, 
one obtain it, for instance, performing the limit
$V\rightarrow\infty$ on any line $(L/L_0)=\left(V/V_0\right)^a$ with
$\alpha_\sigma /(1+\alpha_\sigma) \leq a < 1$.}

Moreover, the average number of
corrosion front sites per column $\left<I(t_f)\right>/L$ is found to follow 
a power law of the form: 
\begin{equation}
\frac{\left<I(t_f)\right>}{L}\sim\left(\frac{L}{V}\right)^{-\alpha_I}\;\mbox{with}
\; \alpha_I=0.45\pm0.02\,.  \label{N}
\end{equation}

\begin{figure}[]
\centerline{
\psfig{file=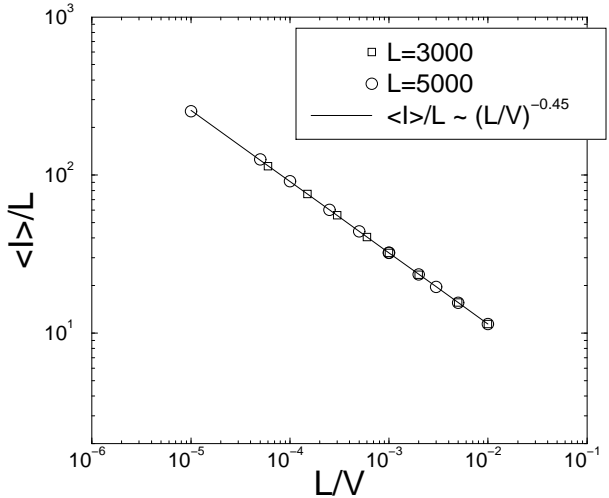,width=8cm}
}
\caption{Scaling behavior of the average number of sites per column in the final
corrosion front.}
\label{fig2a}
\end{figure}

\begin{figure}[bp]
\centerline{
\psfig{file=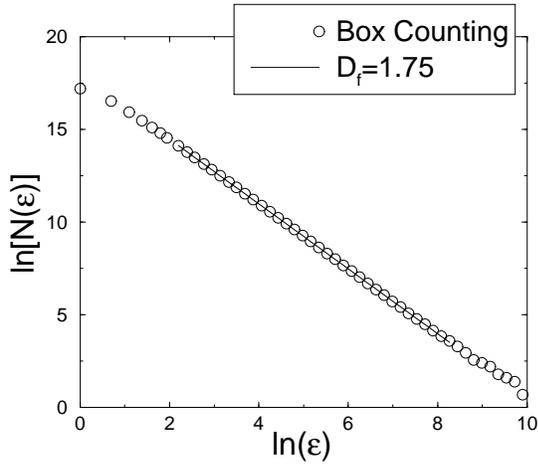,width=7cm}
}
\caption{Box-counting determination of the fractal dimension $D_f$ of
the corrosion front. The value of $D_f=1.753\pm 0.005$ is found
fitting for values of $\epsilon$ ranging from a few lattice distances
to the front width $\sigma$ (in this case $sigma \approx 3000$,
i.e. $\ln(\sigma) \approx 8 $).
\label{dimension}}
\end{figure}

The fractal dimension $D_{f}$ of the corrosion front was measured (up to the scale
$\sigma $) using the box-counting \cite{box-counting} algorithm.
In this way $D_{f}=1.753\pm 0.005$ is found (see Fig.~\ref{dimension}).
Note that it is compatible with the value $7/4$ of GP.

In the early papers  about this etching process \cite{sapo-etch} $D_{f}\simeq 1.62$ was
measured. This different value was due to finite size effects. The present
simulations are almost $400$ times larger than those of those previously reported 
in \cite{sapo-etch} (the largest value of the parameters being $L=3\cdot 10^{4}$ 
and $V=5\cdot 10^{9}$).  
This achievement is important to assert that the exponents characterizing the
final corrosion front belongs to the universality class of Gradient
Percolation.
Whilst $\left<p_{f}\right>$ depends on the lattice geometry, as $p_{c}$ changes, 
the values of the exponents remain the same.

Nevertheless, note that the measured fractal dimension of the corrosion
front can be reduced to $4/3$, if one does not use the right ``dual'' connectivity
criterion introduced above. This is the so-called Grossmann-Aharony
effect in percolation \cite{sapo3,G-H}.
This effect can explain the reduced fractal dimension ($4/3$) measured in the
real corrosion experiments \cite{balazs2}, due to insufficient image resolution.
For example, on the triangular lattice (where the solution etches only f.n.n.),
if the resolution does not distinguish first and second nearest neighbors, the measured
fractal dimension is $4/3$.

\begin{figure}[tbp]
\centerline{
\psfig{file=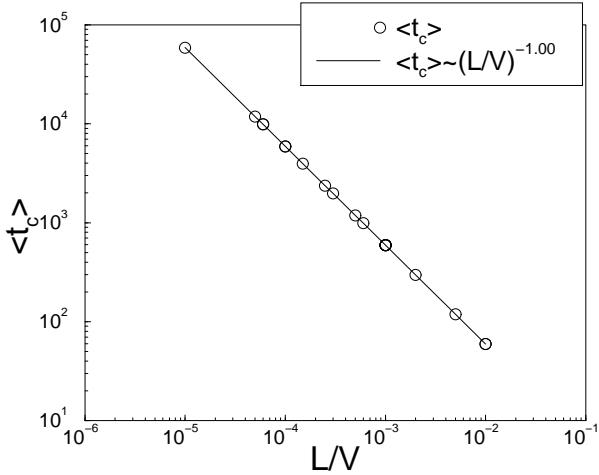,width=8cm}
}
\caption{Scaling behavior of the critical time $t_c$ for which
$p(t_c)=p_c$: $\left<t_c\right>\sim (L/V)^{-\alpha_{t_c}}$ with $\alpha_{t_c}=0.998\pm 0.001$.}
\label{fig-tc}
\end{figure}

\begin{figure}[tbp]
\centerline{
\psfig{file=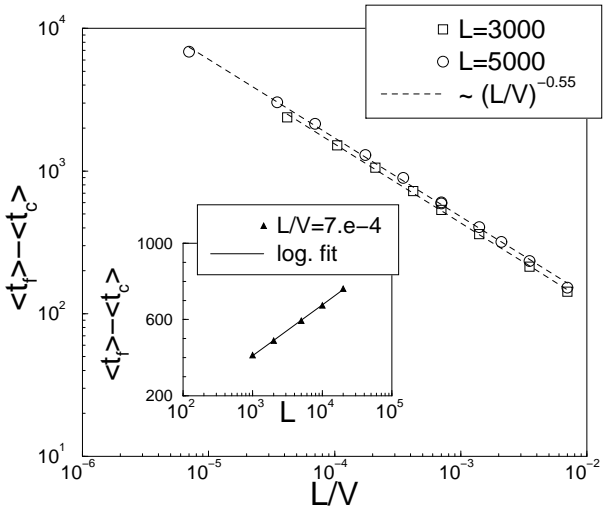,width=8cm,angle=0}
}
\caption{Scaling behavior of $\left<t_f\right>-\left<t_c\right>$ for different
values of $L$. Inset: the dependence on $L$ for $L/V$ fixed is presented.}
\label{fig-tf}
\end{figure}

The average critical time $t_c$, defined by
$p(t_c)=p_c$, and the difference between the arrest time $t_f$ of the dynamics
and $t_c$ itself, are measured for different values of the gradient $L/V$.
For the first one, the following simple behavior is found (see Fig.~\ref{fig-tc}):
\begin{equation}
\left<t_c\right>\sim (L/V)^{-\alpha_{t_c}}\; \mbox{with}\;\alpha_{t_c}=0.998\pm 0.001\,.
\label{tc-eq}
\end{equation} 
As we shall see in the following, this is 
a direct consequence of the linear properties of the smooth dynamical regime (Eq.~\ref{simplest}).
Finally, for $t_f$ one has (see Fig.~\ref{fig-tf})
\begin{equation}
\left<t_f\right>-\left<t_c\right>\sim (L/V)^{-\alpha_{t_f}} \;
\mbox{with}\;\alpha_{t_f} \approx 0.55\,.
\label{tf-eq}
\end{equation}
However, for $\left<t_f\right>-\left<t_c\right>$, a further dependence on $L$ is obtained
(see the inset of Fig.~\ref{fig-tf}). In particular, changing $L$ with $L/V$ fixed,
the quantity $\left<t_f\right>-\left<t_c\right>$ is found to depend linearly on $\ln L$. 
This behavior is connected to the ``extremal'' nature of $t_f$ and is not
studied here \cite{extremal}.

\subsection{Scaling relations}

The exponents $\alpha_{\sigma}$, $\alpha_p$, $\alpha_I$, and $D_f$ are not independent.
At first, note that, within the present numerical precision, 
\begin{equation}
\alpha_{\sigma}=\frac 1{D_f}
\label{sigma-df}
\end{equation}
as in GP \cite{sapo2}.

Identifying the width $\sigma$ with the horizontal correlation length,  
the average number within a correlated region scales as
$\sigma^{D_f}$ because of the fractality on smaller scales. Since the
horizontal size of the solid is $L$, the average number of distinct correlated regions 
will be $L/\sigma$. 
Consequently, one can write: 
\[
\left<I(t_f)\right>\sim \frac{L}{\sigma}\sigma^{D_f}, 
\]
which implies 
\begin{equation}
\frac{\left<I(t_f)\right>}{L}\sim \sigma^{D_f-1}\sim \left(\frac{L}{V}
\right)^{-\alpha_{\sigma}(D_f-1)}. 
\label{I-df}
\end{equation}
From Eqs.~\ref{N},~\ref{sigma-df} and \ref{I-df} one then obtains the following scaling relation: 
\begin{equation}
\alpha_I=\alpha_{\sigma}(D_f-1)=\frac{D_f-1}{D_f},  \label{scaling-N}
\end{equation}
which is consistent with the measurement of $\alpha_I$ in Eq.~\ref{N}. 

Exploiting the analogy between $L/V$ and the gradient $\nabla p$ in 
GP, another interesting relation among exponents can be derived. 
From the relation $\Delta p\equiv p_{c}-p_{f}\approx \nabla p\cdot
\sigma $, one gets: 
\begin{equation}
\alpha_p =1-\alpha_{\sigma} =\frac{D_{f}-1}{D_{f}} 
\label{p-sigma}
\end{equation}
 Note that this implies $\alpha_p =\alpha_I $ in $d=2$. In fact the assumption
that the number of different correlated regions scales as $L/\sigma$ is valid only in $d=2$.

\section{Dynamical equations and theoretical results.}

In this section we present an analytical derivation of the dynamical evolution
of $p(t)$ and the distribution of the surface resistances. This
time-dependent distribution characterizes the evolution of ``hardening'' properties
of the surface.
To this
aim, the histogram $h(r,t)$ is introduced. 
The quantity $h(r,t)dr$ measures the number of global surface sites
with random resistance in the interval $[r,r+dr]$ at time $t$.  
By definition, the number of sites in the global surface $G(t)$ 
is simply the total integral of the histogram: 
\begin{equation}
G(t)=\int_{0}^{1}h(r,t)dr\,.  
\label{eq3}
\end{equation}
On the other side, the number of surface sites being corroded at time $t$ by the solution will be: 
\begin{equation}
n(t)=\int_{0}^{p(t)}h(r,t)dr\,,  
\label{eq4}
\end{equation}
as $n(t)$ is the number of sites in the global surface with $r<p(t)$.
Note that Eq.~\ref{eq4} links $h(r,t)$ directly to $p(t)$ through Eq.~\ref{eq2},
which can then be rewritten: 
\begin{equation}
p(t+1)=p(t)-\frac{\int_{0}^{p(t)}h(r,t)dr}{V}  
\label{eq7}
\end{equation}
Let us call $m(t)$ the number of active sites at time $t+1$: i.e. 
the new sites entering the global surface 
as a consequence of the corrosion of the set of $n(t)$ sites. 
Then the set $m(t)$ is the {\em active zone} at time $t+1$.
One can define $\omega (t)=m(t)/n(t)$. Therefore $\omega (t)$ is the
number of new active sites per etched site at time $t$.
As shown below, the quantity $\omega(t)$ is the fundamental parameter
relating the ``geometry'' to the ``chemistry'' of the system at time $t$. 
At each time-step one can write 
\begin{equation}
G(t+1)=G(t)-n(t)+m(t)\,,
\label{g-m}
\end{equation}
or, using both Eq.~\ref{eq4} and the definition of $\omega(t)$: 
\begin{equation}
G(t+1)=G(t)+(\omega(t)-1)\int_{0}^{p(t)}h(r,t)dr\,. 
\label{eq5}
\end{equation}
Considering only sites in $[r,r+dr]$, one can write: 
\begin{eqnarray}
h(r,t+1) &=&h(r,t)-h(r,t)\theta (p(t)-r)+  \nonumber \\
&&+\omega (t)\int_{0}^{p(t)}h(r^{\prime },t)dr^{\prime }\;,  \label{eq6}
\end{eqnarray}
where $\theta(x)$ is the Heavyside step-function. In Eq.~\ref{eq6}, the
second term in the right-hand side represents the number of sites etched at time $t$ 
(a surface site $i$ is etched with probability $1$ if $r_{i}<p(t)$~).
The third term is the contribution to $h(r,t)$ due to the new active zone.
It is based on the fact that each new site has completely
random resistance to etching; the probability that it belongs to the
interval $[r,r+dr]$ is simply $dr$ (as $\pi_0(r)=1$). 
In principle, knowing the behavior of $\omega (t)$, one can solve the system
given by Eqs.~\ref{eq7} and \ref{eq6},
characterizing in this way the dynamical evolution of the corrosion
power and of the resistance of the solid surface.

Before going on with calculations, it is important to observe that $h(r,0)=L$ 
~$\forall r\!\in\! [0,1]$ (as $\pi_{0}(r)=1$ and the initial surface
is a layer of sites of length $L$). On the other hand for $r<p(t)$, Eq.~\ref{eq6}
reduces to
\begin{equation}
h(r,t+1)=\omega (t)\int_{0}^{p(t)}h(r^{\prime },t)dr^{\prime }\;.
\label{h-red}
\end{equation}
Eq.~\ref{h-red} and the initial condition $h(r,0)=L$ imply that at each time
for $r<p(t-1)$ $h(r,t)$ is independent on $r$ and can be written as: 
\begin{equation}
h(r,t)=L~\prod_{t^{\prime }=0}^{t-1}(\omega (t^{\prime })p(t^{\prime
}))~~~\mbox{for}~~r<p(t-1)  \label{eq8}
\end{equation}
Using this expression in Eq.~\ref{eq7}, the following equation
is obtained: 
\begin{equation}
p(t+1)=p(t)\left[ 1-\frac{L}{V}\prod_{t^{\prime }=0}^{t-1}(\omega (t^{\prime
})p(t^{\prime }))\right] ~.  \label{eq9}
\end{equation}
Eq.~\ref{eq9} makes evident the strong dynamical link between the geometry
($\omega(t)$) and the chemistry ($p(t)$) of the system.

In order to examine the calculations further, it is necessary
to make some hypothesis on the behavior of $\omega (t)$.

As previously mentioned, the dynamical evolutions can be divided into two regimes: 
\begin{enumerate}
\item  a first {\em smooth regime},  which is referred to the
time scale at which $p(t)$ is larger than $p_{c}$;

\item  a second {\em critical regime}, which is referred to
the time scale at which $p(t)\simeq p_{c}$.
\end{enumerate}

This partition of the dynamics into two regimes is directly 
connected to percolation theory \cite{stauffer}, as shown below
 
\subsection{Smooth regime}

If one considers all the lattice sites with $r<p(t)$ for $p(t)>p_{c}$, they 
form both a set of a few finite size clusters and an infinite
percolating and homogeneous (not fractal) cluster \cite{stauffer}. 
Consequently, the intersection
between the global solid-solution surface and this set is
made of a large number of sites. The larger $p(t)$, the larger the
intersection.
This intersection is nothing else but the set of $n(t)$ sites to
be dissolved at that time-step.

Since $n(t)\gg 1$ (and then $m(t-1)>n(t)\gg 1$ also), one can use 
the law of large numbers to relate $n(t)$ to $m(t-1)$: 
\begin{equation}
n(t)=p(t)m(t-1)\,.
\label{n-p}
\end{equation} 
For the same reason one expects small fluctuations around these values. 
From Eq.~\ref{n-p} and the definition of $\omega (t)$, one obtains 
\begin{equation}
m(t)=\omega(t)p(t)m(t-1)\,.
\label{m-m}
\end{equation}
Because of the percolation properties for $p>p_c$, which are related to
the previous argument, one expects
\[
\left| \frac{m(t)-m(t-1)}{m(t-1)}\right| \ll 1~.
\]
Hence the relation: 
\begin{equation}
\omega (t)\simeq \frac{1}{p(t)}~.  
\label{linear}
\end{equation}
This equation introduces an important relationship between
the fundamental ``chemical'' parameter $p(t)$ and the ``geometrical'' and ``dynamical'' 
parameter $\omega(t)$. 
Inserting Eq.~\ref{linear} in Eq.~\ref{eq9}, one recovers Eq.~\ref{simplest}, 
which can be rewritten as: 
\begin{equation}
p(t)=p(0)\exp \left( -\frac{t}{\tau }\right)  \label{eq10}
\end{equation}
with $\tau =-1/\ln (1-L/V)\simeq V/L$ ~($L\gg V$). 
From Eq.~\ref{eq10} one has ~$t_c\simeq V/L \ln(p_0/p_c)\sim (L/V)^{-1}$.
This confirms
the numerical result found in Fig.~\ref{fig-tc} and expressed by Eq.~\ref{tc-eq}.
Note that the behavior given by Eq.~\ref{eq10} is independent on the space 
dimension and on the coordination number of the lattice.
This behavior is however valid only up to the time at which $p(t)\simeq p_{c}$.
After this time the hypothesis to deduce Eq.~\ref{eq10}, and in particular the
possibility of using the law of large numbers, breaks
because of the geometrical constraints given by the percolation properties
of random numbers on a lattice.

Using Eq.~\ref{eq10}, one can derive rigorously the shape of $h(r,t)$ at any time-step $t$
or of its normalized version $\phi(r,t)$ (i.e. $\int_0^1 dr \,\phi(r,t)=1$).
$\phi(r,t)$ is obtained by dividing $h(r,t)$ by $G(t)$.
Technical calculations are reported in Appendix B. 
We provide here directly the result:
\begin{eqnarray} 
&&\phi(r,t)\simeq\phi_1(t)\times\nonumber 
\\ 
&&\times\left\{ 
\begin{array}{ll} 
\frac{1}{t} &\mbox{for}\;r\le p(t)\\ 
1-\frac{\tau}{t}(\ln p(0)-\ln r)&\mbox{for}\;p(t)\le r\le p(0)\\ 
1& \mbox{for}\;r\ge p(0) 
\end{array} 
\right.\label{phi2-bis} 
\end{eqnarray} 
where 
\begin{eqnarray}
[\phi_1(t)]^{-1}&=&1+(p(0)-p(t))/\ln(p(t)/p(0))=\nonumber\\
&=&1-\frac\tau t p(0)(1-\exp(-t/\tau))\,.
\end{eqnarray}
Eq.~\ref{phi2-bis} can be rewritten, in term of $p(t)$ instead of $t$, as follows:
\begin{eqnarray} 
&&\phi(r,t)\simeq\phi_1(t)\times\nonumber 
\\ 
&&\times\left\{ 
\begin{array}{ll} 
\frac{1}{t} &\mbox{for}\;r\le p(t)\\
1-\frac{\log(r/p(0))}{\log(p(t)/p(0))}&\mbox{for}\;p(t)\le r\le p(0)\\ 
1& \mbox{for}\;r\ge p(0) 
\end{array} 
\right.\label{phi2-tris} 
\end{eqnarray}

\subsection{Critical regime}

For $p(t)=p_{c}$, one has (in the limit $L\rightarrow\infty$)
a marginal critical case in which the set of lattice sites with $r<p(t)$
form finite size clusters of any size and an infinite fractal percolating cluster. 
Finally for $p(t)<p_{c}$, 
the set of lattice sites with $r<p(t)$ forms only finite size clusters.
For this reason, even if at such time $t$ the intersection between the global solid surface and
the set of lattice sites with $r<p(t)$ is not empty, it becomes depleted after a 
finite number of time-steps. The average number of time-steps after which the dynamics stops 
will be a function of the system parameters $L$ and $V$ (Eq.~\ref{tf-eq}).
At this time the corrosion dynamics stops because the
intersection between the global surface and the set of lattice sites with $r<p(t)$
is empty. This explains why the final corrosion front is fractal with a 
fractal dimension $D_f$ and a characteristic size (thickness) $\sigma$. 
$D_f=7/4$ is the hull fractal dimension of the finite clusters formed by the lattice
sites with $r< p(t_f)$ and $\sigma$ is the characteristic size of these
clusters. Finally, the same argument
explains why each exponent, characterizing the above introduced scaling relations 
(apart from those about $t_c$ and $t_f$), is directly connected to the exponents of GP.

From the above argument, 
it is important to note that, if not empty, the active zone at a time $t>t_c$ 
is composed by a small and fluctuating number of sites $m(t)$.
This implies that also  $n(t)$ is small and strongly fluctuating. 
Consequently, the arguments developed in dealing with the smooth time-regime,
based on the law of large numbers and small fluctuations, are no longer valid:
$\omega(t)$ becomes a strongly fluctuating quantity.
These critical fluctuations are related to the
fractal morphology of the critical phase of percolation.
We can say that the arrest of the etching dynamics is due to one of these big fluctuation
of $\omega$ in which no site of the active zone has $r<p(t)$.

\begin{figure}[tbp]
\centerline{
\psfig{file=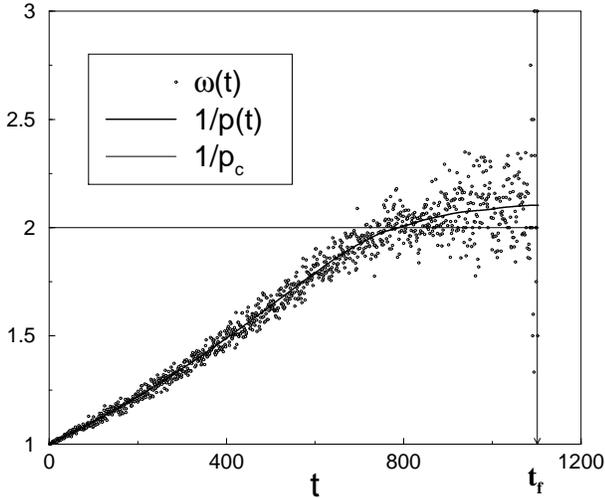,width=8cm}
}
\caption{Comparison between $\omega(t)$ (dots) and $1/p(t)$ (thick line) as
functions of time.}
\label{omega}
\end{figure}

All these features are shown by Fig.~\ref{omega}, where $\omega(t)$ and $1/p(t)$
are shown as functions of time. 
It is important to note that, whereas $\omega(t)$ is a strong fluctuating
quantity in the critical time-regime, $p(t)$ is always smooth.
In fact $p(t)$ can be written (Eq.~\ref{eq2}) as 
\[p(t)=p(0)-\frac{1}{V}\sum_{k=0}^{t-1} n(t)\,.\]
Consequently, $p(t)$ can be seen, apart from prefactors, as the time integral of $n(t)$,
which is a limited function of time and then $p(t)$ is continuous.
Moreover, Fig.~\ref{omega} shows that in the critical time-regime the equality
$\omega(t)\simeq 1/p(t)$ is valid only ``in average'':
\begin{equation}
\overline{\omega(t)}\simeq \frac{1}{p(t)}\,,
\label{om-average}
\end{equation} 
where $\overline \omega(t)$ means the average of $\omega$ over a sufficiently 
large time interval around $t$ in the critical time regime. 
In order to justify the smooth behavior of $p(t)$ in spite the fluctuations of $\omega(t)$,
one can use Eq.~\ref{eq9}.
In the continuous time limit, it can be rewritten as:
\begin{equation}
\frac{dp(t)}{dt}=-\frac{L}{V}p(t)\exp \left[\int_0^t dt'(\ln p(t')+\ln \omega(t'))\right]\,.
\label{eq-p-cont}
\end{equation}
Since the fluctuating $\omega(t)$ appears in the integral term 
of the exponential of the right-hand side of Eq.~\ref{eq-p-cont}, one
expects $p(t)$ to be regular.

In order to understand why one has $\overline{\omega(t)}\simeq 1/p(t)$ also 
in the critical regime, one has to analyze the behavior of time average 
\[\overline{\omega(t)}\equiv \frac{1}{\Delta t}\sum_{t'=t-\Delta t/2}^{t+\Delta t/2}
\omega(t')\,.\]
Using the definition of $\omega(t)$, one can write:
\begin{equation}
\overline{\omega(t)}= \frac{1}{\Delta t}\sum_{t'=t-\Delta t/2}^{t+\Delta t/2}
\left[\frac{m(t'-1)}{n(t')}\,.
\frac{m(t')}{m(t'-1)}\right]
\label{om-average2}
\end{equation}
From Eq.~\ref{n-p} one has directly 
\[\sum_{t'=t-\Delta t/2'}^{t+\Delta t/2} [m(t'-1)/n(t')]=1/p(t).\] 
From this observation and supposing that the quantity $m(t')/m(t'-1)$ oscillates symmetrically 
around $1$, one expects Eq.~\ref{om-average}.
Before proceeding, it is worth to note that the variation of $p(t)$ during the critical time-regime,
as shown by Eq.~\ref{delta-p}, is very small. From this one has approximatively:
\begin{equation}
\overline{\omega(t)}\simeq 1/p_c
\label{om-crit}
\end{equation}
Eqs.~\ref{om-average} and \ref{om-crit} are very important because they provide both
a ``physical'' and a ``geometrical'' meaning to the critical threshold of percolation in a 
given lattice (an analogous relation for Invasion Percolation was found in
\cite{matteo,rts}). 

In order to clarify the geometrical effect on $\omega(t)$, 
it is important to observe that, following Eq.~\ref{linear} and Eq.~\ref{eq10},
$\omega(t)$ would increase to infinity. Clearly this is not
possible in a finite dimensional lattice with a finite coordination
number. For instance in a $2d$ site square lattice with f.n.n.
 connection $\omega$ must be always be smaller than
$3$. Moreover, as seen above, the percolation theory introduce a stronger constraint
forbidding  $p(t)$ to go  well below $p_c$. \footnote{Note that
$p_c$ is always larger than the inverse of the coordination number of
the lattice~\cite{stauffer}. They are equal only in the Bethe lattice without loops.}
Consequently, in order to use Eq.~\ref{eq9} to predict the behavior of $p(t)$, one
has to take in account both the behavior given by Eq.~\ref{om-average} and these 
geometrical constraints.  
In order to show this, we have made the approximation $\omega(t)=1/p_{sim}(t)$ 
in Eq.~\ref{eq9}, where $p_{sim}(t)$ is the
simulation outcome for $p(t)$ (and then it includes automatically the
geometrical constraint). The solution $p(t)$ obtained by solving numerically Eq.~\ref{eq9}
is very near to $p_{sim}(t)$ itself.

Because of the strong fluctuations, the purely analytical study of this critical regime 
is difficult.
For this reason we developed only an approximated mean field approach 
by imposing only the geometrical constraint with the following simple approximation:
\begin{equation}
\omega(t)=1/p_c  
\label{critical}
\end{equation}
in this critical time-regime.
This
is a kind of mean field approximation as the
fluctuations of $\omega(t)$ are neglected. Inserting the relation
\ref{critical} in Eq.~\ref{eq9}, one can write:
\begin{equation}
p(t+1)=p(t)\left[1-\frac{L}{V}p_c^{-t}
\prod_{t^{\prime}=0}^{t-1}p(t^{\prime})\right]  \label{eq13}
\end{equation}
with the initial condition $p(t_c)=p_c$ and $t>t_c$. 
In order to solve Eq.~\ref{eq13}, one can consider the continuous limit  which
is equivalent to
take $\omega(t)=1/p_c$ in Eq.~\ref{eq-p-cont}:
\begin{equation}
\frac{dp(t)}{dt}=-\frac{L}{V}p(t)\exp \left[\int_0^t dt'(\ln \frac{p(t')}{p_c})\right]\,.
\label{eq-p-pc}
\end{equation}
This equation can be solved exactly, and this
solution is well 
approximated by: 
\begin{equation}
p_c-p(t)\simeq p_c \sqrt{\frac{2L}{V}}\frac{\exp\left(\sqrt{\frac{2L}{V}}(t-
t_c)\right)-1} {\exp\left(\sqrt{\frac{2L}{V}}(t-t_c)\right)+1}  \label{eq15}
\end{equation}
where $t_c$ is the time at which $p(t)=p_c$. From Eq.~\ref{eq15} we can see that 
in the $t\rightarrow\infty$ limit one has 
\begin{equation}
p_c-p_f\sim \sqrt{\frac{L}{V}}~.  \label{eq16}
\end{equation}
In spite of the rough approximation, we obtain a good approximation of the numerical  
result of Eq.~\ref{delta-p} (see also Fig.~\ref{fig2}), then
this ``mean field'' result provides a good approximation.

Moreover, since the time constant of Eq.~\ref{eq15} is $\tau^{\prime}\sim 
\sqrt{\frac{V}{L}}$, it is argued that the average time $t_f$ at which the
dynamics spontaneously stop, obeys the following scaling law: 
\[
t_f-t_c \sim \tau^{\prime}\sim \sqrt{\frac{V}{L}}. 
\]
Also, this is a reasonable approximation with respect to the simulation
behavior (Eq.~\ref{tf-eq} and Fig.~\ref{fig-tf}). 
We can see that Eq.~\ref{eq16}
can be interpreted as the product of the $\Delta p$ in single time steps 
( $\Delta p\sim L/V$) multiplied by the average number of time-steps necessary to
stop the dynamics $t_f-t_c \sim\sqrt{\frac{V}{L}}$.

\section{Final histograms and surface hardening}

\begin{figure}[tbp]
\centerline{
\psfig{file=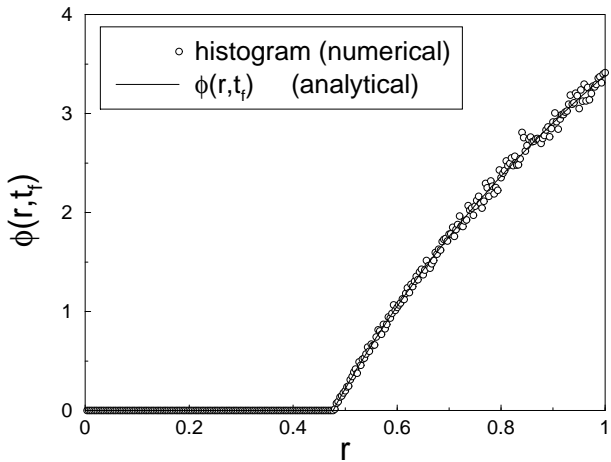,width=8cm}
}
\caption{Normalized histogram of site resistances belonging to the final global surface.
Circles represent the result of numerical simulations and
the solid line the theoretical prediction detailed 
in Appendix B.}
\label{histogram}
\end{figure}

The surface hardening can be described by considering both the histogram of
the global surface and that of the corrosion front.
One first observes that the number of corroded sites in
the critical regime is much smaller
than the number of sites which have been etched in the smooth regime.
This means that the global surface is dominated by sites belonging to finite 
clusters. 
As a consequence, $\phi(r,t_f)$ is well approximated by the linear regime behavior
given in 
Eq.~\ref{phi2} with  $t=t_f$ and $p(t)=p_f$ (see Fig.~\ref{histogram}).
One observes good agreement between theory and numerics.

The global histogram describes the hardening phenomenon
of the global surface which includes finite size clusters detached 
by the etching process at various time-steps of the dynamics.
The increasing behavior of $\phi(r,t_f)$ for $r>p_c$ is 
due to the fact that the majority of surface sites belonging to finite 
islands have been discovered at intermediate time-steps when
$p(t)>p_c$. This implies that their resistance is well above $p_c$.

\begin{figure}[tbp]
\centerline{
\psfig{file=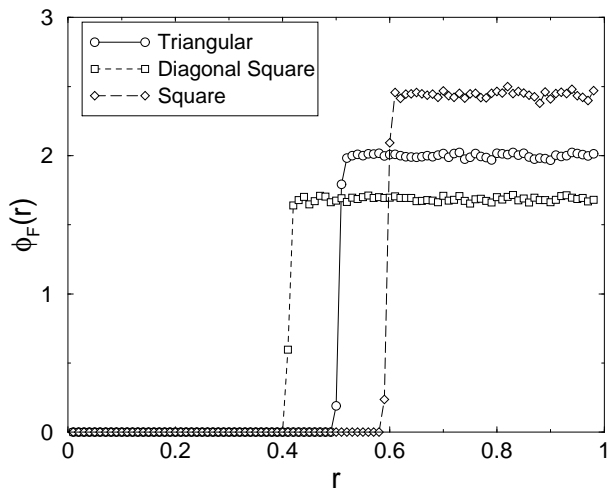,width=8cm}
}
\caption{Normalized histogram of the corrosion front
resistances. One can notice that simple step functions are found for 
any three different lattices studied. The discontinuities occur 
very near the percolation thresholds.}
\label{front-hist}
\end{figure}

The hardening of the corrosion front is described
by the normalized distribution $\phi_F(r)$ of it site resistances.
It has been measured by numerical simulation for several types of 
lattices.
The numerical results are shown in Fig.~\ref{front-hist}   

As discussed above, the final front is more resistant to etching
than the original native surface. This is shown in Fig.~\ref{front-hist}, 
giving the histogram of the front resistances. In first approximation it is a step
function around $r=p_c$. This confirms the hypothesis, derived by GP, 
that the final corrosion front corresponds to the hull of percolation 
clusters with $p=p_f$.

This effect could possibly be used practically in waste
management problems. Consider for instance a system 
containing dangerous compounds with random resistances to etchants present
in the environment.
If in natural circumstances it comes in contact with etchants,
even with a weak etching power $p_0<p_c$, then dangerous materials can be diffused in 
the environment. 

But one can think to apply to the system
a previous etching treatment with $p_0^{'}>p_c$, in an artificially controlled situation.
In this case  the final surface contains only sites with $r>p_c$.
Then the treated surface will resist 
forever to any further natural attack with $p<p_c$, with no danger of leaks in nature. 
This could be called a random ``Darwinian'' selection of a strong surface : 
once selected with a finite solution with
$p >p_c$ the surface resists for ever to further etching with $p_0<p_c$ whatever the 
volume of the solution.

\section{Discussion}
Several properties and limitations of this model should be
discussed. This depends on the nature of what has been described here as a 
``site''. One can think of a site as being an atom or a small group of atoms.
For example, if one considers a random solid (like a glass) the
different local random environments will cause random local rates for atomic
dissolution rather than random probabilities. The possible application of the
above model is then restricted to situations where the choice of
suitable time intervals makes it possible to separate ``very resistant''
and ``very weak'' sites. This implies a transposition of the distribution of
local rates into a distribution of sites resistances.

But a site could also be a semi-macroscopic entity like a small crystallite
protected by a randomly resistive surface.    
In the case of corrosion experiments by Bal\`azs \cite{exp-etc}
it is believed that randomness may be attributed to the random nature of the oxide layer
which spontaneously grows on previously etched aluminum crystallites. The
disorder studied herein occurs if the random resistances to
etching appear just after oxidation of newly discovered
crystallites. Although the disorder appears dynamically, once created
it is quenched. 

It is worth to note that a different kind of process would lead to the
same description.
Consider for example a case where crystallite
oxidation and corrosion are two possible processes in competition when
a site is uncovered.
If the
probability $p$ associated to the corrosion is proportional to the
global etchant concentration, then  a probability
$1-p$ should be associated to oxidation (and then passivation). 
In order to decide the etching or passivation of a given site $i$, a
random number $r_i$ is thrown. If $r_i$ is smaller then $p$ the site
is corroded, otherwise is defintively passivated.
The $r_i$ numbers define a stochastic process which would give the
same dynamical behavior.  Of course the hardening properties would be
different in that case.
From a statistical point of view, it
means simply that we can equivalently formulate our model as a
deterministic dynamics with quenched disorder or as a stochastic
dynamics without quenched randomness.

\section{Conclusions}

In this paper we have discussed several aspects of a simple model for
the etching of a two-dimensional disordered system by a finite volume
of corroding solution.  This has been done both theoretically and
verified numerically .  The dynamics correspond to the disappearance
of weak surface sites which at the same time uncovers new sites. As
the etching process consumes the etchant, the etching power of the
solution decreases and the surface resistance increases to the point
where the process stops spontaneously. One obtains a kind of
``equilibrium'' or static situation in which the dynamics is
stopped. This static state is characterized by the fact that the
surviving interface sites have a resistance to the etching which is
larger than the final value $p_f$ of the solution etching power which
is on the order of the percolation threshold $p_c$.

An analytical description of the time behavior of the solution etching
power $p(t)$ and of the distribution of resistances on the total
interface has been introduced. This analytical approach indicates why
and how the dynamics can be divided into two regimes. The first
initial period corresponds to a classical, or superuniversal,
regime. It can be described with precision by a mean field
approximation. The second regime is a critical regime related to
percolation criticality. The final connected interface is constituted
by a collection of fractal interfaces up to a certain characteristic
depth or scale $\sigma$.  The fractal dimension is found to be very
close to $D_f=7/4$. The difference $p_c-p_f$ between $p_c$ and the
final etching power $p_f$, and the width $\sigma$ are both linked to
the geometrical and external parameters characterizing the system via
simple scaling relations. These properties can be simply explained by
relating the model to the Gradient Percolation model: identifying the
ratio $L/V$ between the size of the solid and the volume of the
solution with the gradient $\nabla p$ which characterizes the scaling
properties of GP. After this identification, it has been shown that
our etching model belongs to the GP universality class, and that the
exponents can be explained through percolation theory.

An important result of this approach is the identification of the
meaning of $p(t)$ as the inverse of the mean number of new interface
sites uncovered by each etched site. This identification in particular
is very important in relation to the static final situation in which
$p(t)\simeq p_c$, as it provides the physical meaning of the
percolation critical threshold.

Several further developments of these studies can be suggested.  
The statistics of other observable quantities, as
the arrest time of the process $t_f$ or the maximal depth attained by
the solution, can be studied and related to known results of
asymptotic extreme theory \cite{extremal}. In particular, such
statistics determine the probability of ``chemical'' fracture of a
finite solid submitted to an etching process. Furthermore, the
distribution of the debris produced by the etching process, can be
regarded as a ``chemical'' fragmentation process. 

Also, the stability of the final (harder) interface with 
respect to external perturbations (as for example, fluctuation 
of the etching power $p_f$)
should be of interests.

Authors would like to acknowledge M. Filoche and M. Dejmek for 
interesting discussions and a critical reading of the manuscript.
This work has been supported by the European Community TMR Network
``Fractal Structures and Self-Organization" ERBFMRXCT980183.
The ``Laboratoire de Physique de la Mati\`ere Condens\'ee de 
l'Ecole Polytechnique''
and the ``Centre de Math\'ematiques et de leurs Applications de l'Ecole
Normale Superieure de Cachan'' are `` Unit\'es mixtes de recherches du CNRS''
under the respective numbers 7453 and 8536. 
\appendix
\section{Thermodynamic limit}

In light of what was written about the ``phase'' diagram, 
one can discuss the thermodynamic limit.
Let us start with a couple of external parameters $(L_0, V_0)$ in the
strong gradient ``phase'' for which  $\sigma(L_0,V_0)\ll L_0$, and negligible boundary effects.   
If one changes $L$ (horizontal size of the solid) and $V$ (volume of the solution) 
by satisfying the relation: 
\[\frac{L}{L_0}=\left(\frac{V}{V_0}\right)^{\frac{\alpha_\sigma}{1+\alpha_\sigma}}\,,\]
the new system remains in the strong gradient ``phase'' for any value of $V$.
The final corrosion front obeys the relation:
\[\frac{\sigma(L,V)}{L}=\frac{\sigma(L_0,V_0)}{L_0} \,,\]
i.e. the new system is geometrically ``similar'' to the old one.

Let us now study what happens changing $(L,V)$  following the relation
\begin{equation}
\frac{L}{L_0}=\left(\frac{V}{V_0}\right)^a
\label{thermo}
\end{equation}
with $a\neq \alpha_\sigma /(1+\alpha_\sigma)$.
\begin{itemize}
\item If $a > 1$ the system stays in the strong gradient phase for any value of $V$, 
but $\sigma(L,V)$ is a
decreasing function of $V$ and in the thermodynamic limit ($V\rightarrow\infty$), the final
corrosion front is ``microscopically'' flat; 
\item If $\alpha_\sigma /(1+\alpha_\sigma) < a \leq 1$ the system stays in the  strong 
gradient phase.
Moreover $\sigma(L,V)$ increases with $V$ (it is infinite in the thermodynamic limit),
but $\sigma(L,V)/L < \sigma(L_0,V_0)/L_0$. Hence the new corrosion front is not
``similar'' to the old one, and in the limit $V\rightarrow\infty$ it becomes
``macroscopically'' flat;
\item If $a < \alpha_\sigma /(1+\alpha_\sigma)$, $\sigma(L,V)/L$ increases with $V$.
However there will be a marginal (or ``critical'') value $V_c$ of the volume for which
the geometric correlations reach their maximum possible value: $\sigma(L,V)\simeq L$.\
Increasing $V$ further, following Eq.~\ref{thermo}, the system enters the weak gradient phase 
dominated by boundary effects.
\end{itemize}

\section{The ``global'' histogram}
In the smooth regime, at each time step $t$, the number of surface sites which
are created is $L$, and the number of dissolved sites is $p(t)L$.
Consequently, at each time-step, the total number of surface sites increases 
by $(1-p(t))L$. More precisely, we can see that these last sites affect only
the high $r$ part of the histogram. In fact,
using Eq.~\ref{eq8} in Eq.~\ref{eq6}, one can write the explicit
form of  $h(r,t)$ for any time-step in the smooth regime: 
\begin{eqnarray}
&&h(r,t)=\\
&&=\left\{
\begin{array}{lllll}\nonumber
\!L & \mbox{for}~r<p(t-1) &  &  &  \\ 
\!...... &  &  &  &  \\ 
\!(t-t^{\prime })L & \mbox{for}~p(t^{\prime }+1)<r\le p(t^{\prime })\;;~t^{\prime
}\le t-2 &  &  &  \\ 
\!...... &  &  &  &  \\ 
\!(t+1)L & \mbox{for}~r>p(0) &  &  & 
\end{array}
\right.  \label{eq11}
\end{eqnarray}
Note that $h(r,t)$ is, at any time, a non-decreasing
multi-step function of $r$. 
Each step differs from the previous one by height $L$.

It can be written also: 
\begin{eqnarray}
&&h(r,t)=L~\theta \left( p(t-1)-r\right) +\sum_{t^{\prime }=0}^{t-2}\left[
(t-t^{\prime })L\times \right. \\
&&\left. \times \theta \left( r-p(t^{\prime }+1)\right) \theta \left(
p(t^{\prime })-r\right) \right] +(t+1)L~\theta \left( r-p(0)\right) ~. 
\nonumber
\end{eqnarray}
In a similar way, the explicit
function $G(t)$ is obtained by inserting Eq.~\ref{eq8} and Eq.~\ref{eq10} in Eq.~\ref{eq5}:
\begin{equation}
G(t)=
L\left[ t+1-p(0)\frac{1-\exp (-\frac{t}{\tau })}{1-\exp (-\frac{1}{\tau })}\right]  
\label{eq12}
\end{equation}
In order to obtain the normalized histogram $\phi(r,t)$ one has to divide 
Eq.~\ref{eq11} by Eq.~\ref{eq12}.
Because of the step-like shape of Eq.~\ref{eq11}, $\phi(r,t)$ will also be step-like.

A more physical derivation of a smooth function 
interpolating the step-like $\phi(r,t)$
can be obtained under the same approximate phenomenological
approach leading to Eq.~\ref{simplest}.
Under these assumptions, the corrosion front is located at
depth $y$ at the time $t=y$ when the solution has the corrosive power
$p(t=y)$, where $p(t)$ is given by Eq.~\ref{simplest} (or alternatively by Eq.~\ref{eq10}). 
From Eq.~\ref{simplest}, one can deduce that the
solution attains the etching power $p$ at the time $t(p)$ given by
\begin{equation}
t(p)=\frac{\ln(p/p(0))}{\ln(1-L/V)}
\label{t-p}
\end{equation}
when the front is at depth $y=t(p)$.
From this equation, it is possible to infer that a site with random resistance 
$p(0) <r<p(t)$ is etched only if it is located at a depth $y< t(r)$.
In fact, if $y>t(r)$, the site would be reached by the solution
(i.e. checked by the dynamics)
when the etching power $p(t)$ is weaker than its resistance $r$. 
Hence, it is necessary to distinguish three cases, 
in order to write the number of solid sites, with resistance in
$[r,r+dr]$,  belonging to the global surface at time $t$:
\begin{enumerate}
\item All the sites with $r > p(0)$ checked by the
dynamics, resisted to the corrosion, and hence belong to the
surface. Their number is given by $dr$ multiplied by the area spanned
by  the front up to time $t-1$ (included), in addition to such sites
on the corrosion front at time $t$: $(L\,t+L)\,dr$;
\item All the sites with $p(t) < r < p(0)$ checked by the dynamics
before time $t(r)$, have been etched, whereas such sites, checked between $t(r)$ and $t$, 
resisted. The number is given by $dr$ multiplied by the area
spanned by the front between times $t(r)$ and $t-1$ (included), in
addition to the sites on the front at time $t$: $[(L(t-t(r))+L]\,dr$;
\item All the sites with $r < p(t)$ checked by the dynamics, have been
etched.  Only sites with such resistence belonging to the corrosion front at time $t$ 
contribute to the histogram. Their number is $L\,dr$.

\end{enumerate}


One can then write:
\begin{equation}
h(r,t)=\left\{
\begin{array}{ll}
L&\mbox{for}\;r\le p(t)\\
L\left(t-t(r)+1\right)&\mbox{for}\;p(t)\le r\le p(0)\\
L(t+1)&\mbox{for}\;r\ge p(0)
\end{array}
\right.
\label{histo-phen}
\end{equation}
Using the explicit formula for $t(r)$ given by Eq.~\ref{t-p}, with $r$ replacing $p$, 
one has:
\begin{equation}
G(t)=\int_0^1 h(r,t) dr=L\left[ t+1-\tau p(0)\left(1-e^{-\frac t \tau}\right)\right]\,,
\label{g-2}
\end{equation}
where $\tau=(\ln(1/L/V))^{-1}\simeq V/L$.
Note that Eq.~\ref{g-2} differs from the rigorous Eq.~\ref{eq12} only from the
approximation $\exp(-1/\tau)\simeq 1-1/\tau$ which is valid in the present study where
$V/L\gg 1$.
The normalized distribution at all time is $\phi(r,t)\equiv h(r,t)/G(t)$, and for
$t\gg 1$, it can be written as: 
\begin{eqnarray} 
&&\phi(r,t)\simeq\phi_1(t)\times\nonumber 
\\ 
&&\times\left\{ 
\begin{array}{ll} 
\frac{1}{t} &\mbox{for}\;r\le p(t)\\ 
1-\frac{\tau}{t}(\ln p(0)-\ln r)&\mbox{for}\;p(t)\le r\le p(0)\\ 
1& \mbox{for}\;r\ge p(0) 
\end{array} 
\right.\label{phi2} 
\end{eqnarray} 
where 
\begin{eqnarray}
[\phi_1(t)]^{-1}&=&1+(p(0)-p(t))/\ln(p(t)/p(0))=\nonumber\\
&=&1-\frac\tau t p(0)(1-\exp(-t/\tau))\,.
\end{eqnarray}
Eq.~\ref{phi2} is the same as Eq.~\ref{phi2-bis}.


\begin{references}

\bibitem{evans} U. R. Evans, {\em ``The Corrosion and Oxidation of Metals:
Scientific Principles and Practical Applications''}, Arnold, London (1960).

\bibitem{uhlig} H. H. Uhlig, {\em ``Corrosion and Corrosion Control''},
Wiley, New York (1963).

\bibitem{williams} D. E. Williams, R. C. Newman, Q. Song, R. G. Kelly,
Nature, {\bf 350}, 216 (1991).

\bibitem{nagatani} T. Nagatani, Phys. Rev. A, {\bf 45}, 2480 (1992).

\bibitem{meakin} P. Meakin, T. J$\ddot{\mbox{a}}$ssang, J. Feder,
Phys. Rev. E, {\bf 48}, 2906 (1993).

\bibitem{reigada} R. Reigada, F. Sagu\`es, J. M. Costa,
J. Chem. Phys., {\bf 101}, 2329 (1994).

\bibitem{balazs2} L. Bal\'azs, J-F. Gouyet, Physica A,  {\bf 217}, 319-338
(1995).

\bibitem{exp-etc}  L. Bal\'azs, Phys. Rev. E {\bf 54}, 1183 (1996).

\bibitem{sapo-etch}  B. Sapoval, S. B. Santra and Ph. Barboux, Europhys.
Lett., {\bf 41}, 297 (1998). S. B. Santra and B. Sapoval,
Physica A., 266, 160-172 (1999).

\bibitem{sapo2}  B. Sapoval, M. Rosso and J. F. Gouyet, J. Phys. Lett.
(Paris), {\bf 46}, L149 (1985).

\bibitem{sapo-pc} M. Rosso, J. F. Gouyet, B. Sapoval,
Phys. Rev. B, {\bf 32}, 6035 (1985); R. M. Ziff, B. Sapoval,
J. Phys. A, {\bf 19}, L1193 (1986). 

\bibitem{box-counting} K. J. Falconer, {\em ``Fractal Geometry: Mathematical Foundations and
Applications''}, Wiley, New York (1990).

\bibitem{sapo3}  B. Sapoval, M. Rosso and J. F. Gouyet, in {\em ``The Fractal
Approach to Heterogeneous Chemistry''}, edited by D. Avnir (John Wiley and
Sons Ltd., New York, 1989).

\bibitem{G-H}  T. Grossman and A. Aharony, J. Phys. A, {\bf 20}, L1193
(1987).

\bibitem{matteo}  M. Marsili, J. of Stat. Phys., {\bf 77}, 733 (1994); 
A. Gabrielli, R. Cafiero, M. Marsili and L. Pietronero, J. of Stat. Phys., 
{\bf 84}, 889 (1996).

\bibitem{rts}  A. Gabrielli, R. Cafiero, M. Marsili and L. Pietronero,
Europhys. Lett., {\bf 38}, 491 (1997).

\bibitem{duplantier} H. Saleur, and B. Duplantier, Phys. Rev. Lett., {\bf 58}, 2325 (1986).

\bibitem{stauffer} D. Stauffer and A. Aharony, {\em Introduction to Percolation Theory},
2$^{\mbox{nd}}$ ed., Taylor \& Francis Ltd. (1991).

\bibitem{extremal} A. Baldassarri, A. Gabrielli, and B. Sapoval, {\em Statistics of extremal 
events in corrosion dynamics: etching {\bf vs.} fractures}, in preparation.
 
\end{references}
\end{document}